\documentclass[aps,prd,twocolumn,showpacs,groupedaddress,floatfix]{revtex4-1}

\usepackage{graphicx}
\usepackage{dcolumn}
\usepackage{bm}
\usepackage{amsmath}
\usepackage{amssymb}
\usepackage{makecell}
\usepackage{array}
\usepackage{enumitem}
\usepackage[utf8]{inputenc}

\usepackage[]{units}

\newcommand{\mean}[1]{\left\langle #1 \right\rangle}

\newcommand{\Eq}[1]{Eq.~\eqref{#1}}

\newcommand{\Fig}[1]{Fig.~\ref{#1}}

\begin{document}

\title{BKT phase transitions in two-dimensional non-Abelian spin models}

\author{Oleg Borisenko}
\email{oleg@bitp.kiev.ua}
\affiliation{Bogolyubov Institute for Theoretical Physics, National Academy of Sciences of Ukraine, 03680 Kiev, Ukraine}

\author{Volodymyr Chelnokov}
\email{chelnokov@bitp.kiev.ua}
\affiliation{Bogolyubov Institute for Theoretical Physics, National Academy of Sciences of Ukraine, 03680 Kiev, Ukraine}

\author{Francesca Cuteri}
\email{francesca.cuteri@cs.infn.it}
\affiliation{Dipartimento di Fisica dell'Universit\`a della Calabria,
I-87036 Arcavacata di Rende, Cosenza, Italy \\
and INFN - Gruppo collegato di Cosenza, I-87036 Arcavacata di Rende, Cosenza, 
Italy}

\author{Alessandro Papa}
\email{papa@cs.infn.it}
\affiliation{Dipartimento di Fisica dell'Universit\`a della Calabria,
I-87036 Arcavacata di Rende, Cosenza, Italy \\
and INFN - Gruppo collegato di Cosenza, I-87036 Arcavacata di Rende, Cosenza, 
Italy}

\date{\today}           % It is always \today, today,

\begin{abstract}
It is argued that two-dimensional $U(N)$ spin models for any $N$ undergo a
BKT-like phase transition, similarly to the famous $XY$ model. This conclusion 
follows from the Berezinskii-like calculation of the two-point correlation 
function in $U(N)$ models, approximate renormalization group analysis and 
numerical investigations of the $U(2)$ model. It is shown, via Monte Carlo 
simulations, that the universality class of the $U(2)$ model coincides with 
that of the $XY$ model.  
Moreover, preliminary numerical results point out that two-dimensional 
$SU(N)$ spin models with the fundamental and adjoint terms and $N>4$ exhibit 
two phase transitions of BKT type, similarly to $Z(N)$ vector models. 
\end{abstract} 

\pacs{05.70.Fh, 64.60.Cn, 75.40.Cx, 05.10.Ln}

\maketitle

\section{Introduction} 

The existence of the Berezinskii-Kosterlitz-Thouless (BKT) phase transition 
has been established in a number of Abelian models. The most famous example is 
provided by the two-dimensional ($2D$) $XY$ model~\cite{berezin,kosterlitz1,kosterlitz2,xy_proof}. 
It is an infinite-order phase transition characterized by 1) absence of 
singularities in the free energy and all its derivatives, 2) essential 
singularity in the behavior of the correlation length in the vicinity of the 
critical point and 3) power-like decay of the two-point correlation function 
in the massless phase. 

During the four decades following its discovery, a similar phase transition 
has been found and thoroughly studied in many lattice spin and gauge models. 
Below is a list of some of these models, relevant for the present study:
\begin{itemize} 
\item $2D$ $Z(N)$ spin models for $N \geq 5$. Generalized models of 
this type, including vector Potts models, possess two BKT-like phase 
transitions with an intermediate massless 
phase~\cite{Elitzur,savit,kogut,nienhuis,kadanoff,Cardy,Domany,cluster2d,BM10,z5_phys.rev,zN_phys.rev}. 

\item $3D$ $U(1)$ lattice gauge theory (LGT) at finite 
temperature~\cite{parga,svetitsky}. The deconfinement phase transition in this 
model is also of a BKT type. The universality class of the model coincides 
with that of the $2D$ $XY$ model~\cite{beta_szero,u1_isotropic,borisenko,twist_rg,u1_disorder}. 

\item $3D$ $Z(N)$ LGTs at finite temperature for $N \geq 5$ also 
exhibit two BKT phase transitions, which belong to the universality class of 
the $2D$ $Z(N)$ spin models~\cite{3dzn_strcoupl,3dzn_isotropic1,3dzn_full,3dzn_isotropic2}. 

\end{itemize}

This list could be continued to include many anti-ferromagnetic models, 
SOS-type models and some $2D$ models of fermions. 
We would like to emphasize that all known examples where the BKT transition 
takes place in pure systems are restricted to {\em Abelian} spin and gauge models. 
{\em Non-Abelian} systems with weak disorder and, if $D>2$, described, 
{\it e.g.} by the random-field $O(N)$ models, may exhibit a BKT-like phase 
transition which separates the disordered phase from a phase with an infinite 
correlation length~\cite{feldman_1,feldman_2}. 
The main question which we try to answer here is: is there a BKT transition in 
pure {\em non-Abelian} models in dimension $D=2$? If the answer is positive, 
what is its universality class? In the case of the negative answer, the 
intriguing question is: what is the essential physics in non-Abelian models 
which forbids the existence of an infinite-order phase transition? 

To the best of our knowledge, Berezinskii was the first to address this 
question in the context of $2D$ principal chiral models~\cite{berezin}. 
In fact, his answer was positive. However, after discovering the asymptotic 
freedom of non-Abelian $2D$ principal chiral models, the general belief 
is that there are no phase transitions in these models at any finite values of 
the coupling constants (the debate is still open as no rigorous proof of the 
exponential decay of the correlation function at all values of the bare 
coupling constant is available -- see, {\it e.g.},~\cite{seiler}).   

In this paper we discuss non-Abelian spin models where the spins are traces of 
the elements of the non-Abelian group $G$. We study $2D$ lattice models with 
$G=U(N)$, $SU(N)$ and nearest-neighbor interaction. The global symmetry of an
$U(N)$ spin model is $U(1)$. 
According to the Mermin-Wagner theorem, this symmetry cannot be broken 
spontaneously in two dimensions~\cite{mwtheorem} (it should be stressed, 
however, that we are not aware of any proof of this theorem directly applicable 
to the $U(N)$ spin model). Thus, if there is a phase transition in $U(N)$, it 
cannot be accompanied by the breaking of the $U(1)$ symmetry. This is a strong 
indication that, if it exists, such phase transition can be of infinite order, 
for all $N$. 

The phase structure of the $SU(N)$ model, where spins in the action appear only 
in the fundamental representation, is well known for $N=2,3$ and 4, as 
this is the simplest effective model for the Polyakov loops which can be 
calculated in the strong coupling regime of the finite-temperature $3D$ 
$SU(N)$ LGT; all these models exhibit a finite-order phase transition. 
Therefore, in case of $SU(N)$ we are interested here in $N\geq 5$. 
The global symmetry of $SU(N)$ models is $Z(N)$, which can be broken 
spontaneously. However, the symmetry alone is not enough even to get a hint on 
the possible order of the phase transition. For example, even the general 
$Z(5)$ spin model exhibits a complicated phase structure. Depending on the 
relation between the two independent couplings in $Z(5)$, the thermodynamic 
path may cross either only one critical line of the first order phase 
transition or two critical lines of BKT transitions. General $SU(N)$ spin models
possess infinitely many couplings (one for each representation), hence, the 
full phase structure might be quite complicated. 
Various simulations of the finite-temperature $3D$ $SU(N)$ LGT at large $N$ 
show the existence of a first order deconfinement phase transition in the model 
with the fundamental Wilson action (see~\cite{teper} and references therein). 
One expects that the same happens also in the strong coupling regime of the 
model, {\it i.e.} in the $2D$ $SU(N)$ spin model where spins are taken in the 
fundamental representation. As a first step towards general $2D$ $SU(N)$ 
spin models, we consider here the case with both fundamental and adjoint spins, 
in the space of two independent coupling constants.

This paper is organized as follows. In the next section we introduce our 
notations and define $2D$ $U(N)$ and $SU(N)$ spin models. Following 
Berezinskii~\cite{berezin}, we calculate the two-point correlation function in 
$2D$ $U(N)$ models and show that it has a power-like fall-off at large $\beta$. In 
Section~3 we calculate some effective model for both $2D$ $U(N)$ and $SU(N)$ by 
means of the exact integration over original degrees of freedom. 
Using simple combination of the mean field and renormalization group analysis, 
we demonstrate the existence of a BKT phase transition in the $2D$ $U(N)$ model. 
Section~4 outlines some details of our numerical simulations. Section~5 
presents the results of simulations for the $2D$ $U(2)$ spin model, while Section~6 
contains those for $2D$ $SU(N=5,8)$ with fundamental and adjoint terms. 
Summary and Discussions are given in Section~7. 

\section{Two-dimensional spin models}

We work on a $2D$ Euclidean lattice $\Lambda=L^2$, with sites $x=(x_1,x_2)$, 
$x_n\in[0,L-1]$, and denote by $e_n$ the unit vector in the $n$-th direction.
Periodic boundary conditions (BC) are imposed in all directions. Let 
$W(x)\in G$, with $G=U(N)$, $SU(N)$, and $\chi_{\rm f}(W)$, $\chi_{\rm ad}(W)$ 
be the character of $G$ in the fundamental and adjoint representation, 
respectively. 
Consider the following partition function on $\Lambda$, which describes 
the interaction of non-Abelian spins 
\begin{eqnarray}
Z(\beta , N)  \ &=& \ \int \prod_x dW(x)  
\exp  \biggl[ \frac{1}{N^2} \nonumber\\
&\times&\sum_{x,n} \ \biggl( \beta \ \mbox{Re} 
[\chi_{\rm f}(W(x)) \ \chi_{\rm f}(W^*(x+e_n))]  \nonumber   \\ 
&+& \lambda \ \chi_{\rm ad}(W(x)) \ \chi_{\rm ad}(W(x+e_n)) \ \biggr)  \biggr] \nonumber \\
&\equiv& \ \int \prod_x dW(x)\,  e^S \ .
\label{sunpf}
\end{eqnarray}
When $G=U(N)$ we restrict ourself to the fundamental term 
({\it i.e.} we put $\lambda=0$), while for $SU(N)$ we treat the mixed 
fundamental-adjoint action. We normalized both couplings by $1/N^2$, so
that, in the limit $\lambda\to\infty$, the fluctuations of 
$SU(N)$ spins are restricted to the $Z(N)$ subgroup and the fundamental 
part of the action reduces to the action of the $Z(N)$ spin model 
\begin{eqnarray} 
S_{Z(N)} \ &=& \ \beta \ \sum_{x,n} \ \cos\left [ \frac{2 \pi}{N} (k_x - k_{x+e_n}) \right],\nonumber \\
k_x \ &=& \ 0,\cdots, N-1 \ . 
\label{zn_action}
\end{eqnarray}
The trace of a $U(N)$ matrix can be parameterized with the help of $N$ angles, 
{\it e.g.} by taking $W={\rm diag} (e^{i\omega_1}, \cdots, e^{i\omega_N})$.
In this parameterization the part of the action including the fundamental 
characters for both $U(N)$ and $SU(N)$ has the form
\begin{eqnarray}
&&\mbox{Re} [ \chi_{\rm f}(W(x)) \ \chi_{\rm f}(W^{\dagger}(x+e_n)) ] \nonumber \\ &&=\sum_{i,j=1}^N \cos \left  [\omega_i(x)-\omega_j(x+e_n) \right ] \ .
\label{sunaction}
\end{eqnarray} 
For the adjoint character we use the relation $\chi_{\rm ad}(W)=\chi_{\rm f}(W) 
\chi_{\rm f}(W^*)$ (the constant term is omitted). 
The invariant measure for $U(N)$ is given by
\begin{equation}
\int dW  \ = \ \int_0^{2\pi} D(\omega)D^*(\omega) \prod_{k=1}^N 
\frac{d\omega_k}{2 \pi}  \ ,
\label{sunmeasure}
\end{equation}
where
\begin{equation}
D(\omega) \ = \ \prod_{l=2}^{N} \prod_{k=1}^{l-1} \left ( e^{i\omega_k} -  e^{i\omega_l}    \right  ) \ .
\label{Doperator}
\end{equation}
The $SU(N)$ measure coincides with the $U(N)$ one, up to the additional 
constraint 
\begin{equation}
\prod_{k=1}^N \exp [i \omega_k ] = 1 \ ,
\label{suncond}
\end{equation}
which is implemented into the partition function~(\ref{sunpf}) with the help 
of the periodic delta-function
\begin{equation}
 \sum_{n=-\infty}^{\infty}\exp \left [ in\sum_{k=1}^N \omega_k   \right ] \ .
\label{sundelta}
\end{equation}
Due to this constraint, the $SU(N)$ model is invariant only under the global 
discrete shift $\omega_k(x)\to\omega_k(x)+\frac{2\pi n}{N}$ for all $k$ and $x$.
This is just the global $Z(N)$ symmetry. A useful representation for the 
measure, which is used below, reads 
\begin{equation}
D(\omega) D^*(\omega) = \frac{1}{N!} \ \epsilon_{i_1\cdots i_N} 
\epsilon_{j_1\cdots j_N}  \ 
\exp \left [ i \sum_{k=1}^N \ \omega_k (i_k-j_k)  \right ] .
\label{measure_1}
\end{equation}
Here, $\epsilon_{i_1\cdots i_N}$ is antisymmetric tensor and sum over all repeating
indices is understood. 

The partition function~(\ref{sunpf}) can be regarded as the simplest effective 
model for the Polyakov loops which can be derived in the strong coupling region 
of $3D$ LGT at finite temperature. Before going into a complicated analytical 
and numerical study, we would like to present a simple argument which 
shows why the infinite-order phase transition may occur in $U(N)$ models. 
To this end we are going to compute the two-point correlation function given by 
\begin{equation}
 \Gamma_N(\beta , R) \ = \ 
\frac{1}{N^2}\sum_{i,j=1}^N \ \left \langle \ \cos \ \left [ \omega_i(0)-\omega_j(R) \right ] \ 
\right  \rangle \ ,
\label{corrfuncdef}
\end{equation}
where $R$ represents a position on the two dimensional lattice ($R = (x_1,x_2)$).
One expects, at least in the vicinity of the phase transition, that $\Gamma_N(\beta, R)$ 
depends only on $|R|$ and, indeed, this is the case for our approach as will be shown below. 
When $\beta$ is sufficiently small, one can use the conventional strong 
coupling expansion to demonstrate the exponential decay of the correlation 
function. Let us study now the model when $\beta$ is large. Using the 
global symmetry  $\omega_k(x)\to\ - \ \omega_k(x)$, the correlation function 
is presented as 
\begin{eqnarray}
&&\Gamma_N(\beta , R)  \ = \ \frac{1}{Z} \ \frac{1}{N^2}\sum_{s,t=1}^N \ 
\int_0^{2\pi} \prod_x D(\omega_x)D^*(\omega_x)  \nonumber  \\
&&\times \prod_{k=1}^N \frac{d\omega_k(x)}
{2\pi} \exp \left [ \frac{\beta}{N^2} \sum_{x,n} \ \sum_{i,j=1}^N \ \cos \
[\omega_i(x)-\omega_j(x+e_n)]\right.\nonumber\\
&&\left.+ i\sum_x \sum_{k=1}^N h_k(x)\omega_k(x) \right ] 
\ , \label{corrfunc1}
\end{eqnarray}
where we have introduced the sources
\begin{equation}
h_k(x) = \delta_{x,0} \delta_{k,s} - \delta_{x,R} \delta_{k,t}  \ . 
\end{equation}
To compute the correlation function at large $\beta$, we adjust the simple 
argument by Berezinskii~\cite{berezin}. In the context of $U(N)$ models, 
it reduces to replacing the cosine function in the Boltzmann weight by its 
Taylor expansion since, when $\beta$ grows, the system becomes more and more 
ordered. Keeping only the leading quadratic term in the expansion 
and using Eq.~(\ref{measure_1}) for the invariant measure, one gets after
the Gaussian integration (constant term is omitted)
\begin{eqnarray}
\label{corrfunc2}
&&\Gamma_N(\beta \gg 1, R)  \ \approx \ \frac{1}{Z} \ \frac{1}{N^2}
\sum_{s,t=1}^N \ \prod_x \ \frac{1}{N!} \ \epsilon_{i_1\cdots i_N} 
\epsilon_{j_1\cdots j_N}\nonumber  \\ 
&&\times \exp \left [ -\frac{N^2}{4 \beta} \sum_{x,x^\prime} \sum_{k,k^\prime=1}^N 
(h_k(x)+i_k(x)-j_k(x))\right. \nonumber\\
&&\left.G_{k,k^\prime}(x,x^\prime) (h_{k^\prime}(x^\prime)+i_{k^{\prime}}
(x^{\prime})-j_{k^{\prime}}(x^{\prime})) \vphantom{\frac{N^2}{4 \beta}} \right],
\end{eqnarray} 
where the Green function $G$ is given by 
\begin{eqnarray}
G_{k,k^\prime}(x,x^\prime) \ &=& \  \frac{1}{N^2} \ \left ( G_{x,x^\prime} - 
\frac{1}{2} \ \delta_{x,x^\prime} \right )\nonumber\\
&+& \ \frac{1}{2N} \ \delta_{k,k^\prime}  
\delta_{x,x^\prime}  
\label{G_full}
\end{eqnarray}
and $G_{x,x^\prime}$ is the standard $2D$ massless Green function. 
After some algebra we end up with the following expression for the correlation 
function (from now on $R$ denotes its absolute value $R = |x-x^\prime|$, since our approximation does 
not depend on the direction of $x - x^\prime$)
\begin{eqnarray}
\Gamma_N(\beta \gg 1, R)\ \approx \ e^{- \frac{N-1}{4\beta} 
- \frac{D(R)}{2 \beta}} \left( \frac{A(1)}{A(0)} \right )^2 . 
\label{corrfunc_fin}
\end{eqnarray} 
Here, $A(n)$ is a constant which does not depend on $R$
\begin{eqnarray}
&&A(n) \ =\ \frac{1}{N}\sum_{s=1}^N \frac{1}{N!} \epsilon_{i_1\cdots i_N} 
\epsilon_{j_1\cdots j_N}\nonumber\\
&&\times \exp \left [ - \frac{N}{8\beta}\sum_{k=1}^N (i_k-j_k)^2 
- n \frac{N}{4\beta} (i_s-j_s) \right]. 
\label{An}
\end{eqnarray} 
Since $D(R)=G(0)-G(R)\sim \frac{1}{\pi}\ln R$ for large $R$, we conclude that 
the correlation function decays with the power law
\begin{equation}
\Gamma_N(\beta \gg 1 , R)  \ \asymp  \ \frac{\mbox{const}}{R^{\eta}} \ , 
\label{G_largebeta}
\end{equation}
where the index $\eta$ is given by 
\begin{equation} 
\eta \ = \ \frac{1}{2 \pi \beta} \ . 
\label{eta_un}
\end{equation} 
Thus, similarly to the $2D$ $U(1)$ model, $U(N)$ spin models may possess a 
massless phase when $\beta$ is sufficiently large, which is characterized by 
power-like decay of the correlation function. It is also interesting to note 
that the index $\eta$ does not depend on $N$ and coincides with that of the 
$XY$ model. Though far from being rigorous, these simple calculations and the
expression for $\eta$ clearly indicate that all $U(N)$ spin models 
belong to the universality class of the $XY$ model. In particular, there is a 
BKT-like phase transition which separates the phase with the exponential decay 
of the correlation function at small $\beta$ from the massless phase at 
large $\beta$. 

\section{Mean field and renormalization group analysis}

In this section we calculate first an effective model for the partition 
function~(\ref{sunpf}) suitable for the duality transformations. Then we 
proceed to construct a dual representation of the original models. The dual 
formulation is used for the renormalization group (RG) analysis combined with 
the mean field approximation. The latter is supported by the numerical results 
presented in Section~5. 

\subsection{Effective model} 

Since $\chi_{\rm ad}(W)$ can be expressed in terms of $\chi_{\rm f}(W)$ (see
the text after Eq.~(\ref{sunaction})), the action $S$ in the partition 
function~(\ref{sunpf}) depends only on the real and the imaginary parts of the 
fundamental character
\begin{eqnarray} 
\label{real_s}
&&t(x) = \frac{1}{N} \mbox{Re} \chi_{\rm f}\left ( W(x) 
\right ) = \frac{1}{N}\sum_{i=1}^N \cos \omega_i(x) ,  \\
&&s(x) = \frac{1}{N} \mbox{Im} \chi_{\rm f}\left ( W(x) 
\right ) = \frac{1}{N}\sum_{i=1}^N \sin \omega_i(x).
\label{im_t}
\end{eqnarray}
It is convenient to consider the following transformations:
\begin{eqnarray} 
&&\exp \left [ S\left ( \sum_{i=1}^N \ \cos \omega_i,\sum_{i=1}^N \ \sin \omega_i \right )  \right ] \nonumber \\
&&=\int dt ds \ e^{S(N t,N s)} \ \delta\left (N t- \sum_{i=1}^N \ \cos \omega_i \right ) \nonumber\\
&&\times \delta\left (N s- \sum_{i=1}^N \ \sin \omega_i \right ) \ .
\label{transform_pf}
\end{eqnarray}
Making the further change of variables 
\begin{eqnarray} 
\label{r_cos}
t(x) \ = \ \rho(x) \cos\omega(x)  \ ,  \\
s(x) \ = \ \rho(x) \sin\omega(x)  \ ,
\label{r_sin}
\end{eqnarray}
the partition function~(\ref{sunpf}) can be rewritten as 
\begin{eqnarray}
\label{PF_sun_eff}
&&Z(\beta , N) \ = \ \int_0^1 \prod_x \rho(x) d\rho(x)  \ \int_0^{2\pi} 
\prod_x d\omega(x) \\ 
&&\times \  \prod_x \Sigma(\rho(x),\omega(x)) \exp  \left [  \beta \sum_{x,n} \  \rho(x)\rho(x+e_n)\right.\nonumber\\ 
&&\left.\cos (\omega(x)-\omega(x+e_n)) 
+ \lambda N^2 \  \sum_{x,n} \ (\rho(x)\rho(x+e_n))^2  \right  ] \ , \nonumber 
\end{eqnarray}
where $\Sigma(\rho,\omega)$ is the Jacobian of the transformation. For the
$SU(N)$ model, it is given by  
\begin{eqnarray}
&&\Sigma(\rho,\omega)  \ = \  \sum_{n=-\infty}^{\infty} \ \frac{N^2}{N!} 
\ \epsilon_{i_1\cdots i_N} \epsilon_{j_1\cdots j_N}  \ 
\int_0^{2\pi}  \prod_{k=1}^N \ \frac{d\omega_k}{2 \pi}  \nonumber \\ 
&&\times \exp \left [ i \sum_{k=1}^N \ \omega_k (n+i_k-j_k)  \right ] \delta \left (N\rho \cos\omega -  \sum_{k=1}^N \cos\omega_k \right )
\nonumber  \\
&&\times \delta \left (N\rho \sin\omega -  \sum_{k=1}^N \sin\omega_k \right ) \ .
\label{Jacobian}
\end{eqnarray} 
For $U(N)$, only the term with $n=0$ is present. 
Using the integral representation for the deltas in the last expression, 
one can perform all the integrations over $\omega_k$ to get 
\begin{equation}
\Sigma(\rho,\omega)  \ = \  N^2 \sum_{n=-\infty}^{\infty} \ e^{inN\omega} 
\ \Phi_n(\rho) \ ,
\label{Jst}
\end{equation}
where 
\begin{equation}
\Phi_n(\rho)  \ = \ \int_0^{\infty} r \ J_{Nn}(N\rho r) \ \mbox{det} \, J_{n+i-j}(r)
\  dr,
\label{Jnst}
\end{equation}
with  $1\leq i,j \leq N $,  and $J_k(x)$ is the Bessel function. 

\subsection{Dual of spin models} 

The partition function~(\ref{PF_sun_eff}) is well suited for the duality 
transformations. Namely, the effective action for $U(N)$ spin model involving 
$\omega(x)$ variables 
\begin{equation} 
\beta \sum_{x,n} \  \rho(x)\rho(x+e_n) \cos (\omega(x)-\omega(x+e_n)) 
\label{un_action}
\end{equation}
can be considered as the action of the $XY$ model with a space-dependent 
coupling constant. 
Since the Jacobian for $U(N)$ depends only on $\rho(x)$, the conventional dual 
transformations of the $XY$ model can be easily generalized to the present case.
The result on the dual lattice reads 
\begin{eqnarray}
\label{un_dual}
Z_{U(N)}(\beta , N) \, &=& \, \sum_{\{ r(x) \} =-\infty}^{\infty} \, 
\int_0^1 \prod_p \rho(p) d\rho(p)  \, \prod_p \Sigma(\rho(p))\nonumber \\
&&\prod_l \ I_{r(x)-r(x+e_n)}(\rho(p)\rho(p^{\prime}) \beta) \ , 
\end{eqnarray}
where $p$ are plaquettes dual to the original sites, $p,p^{\prime}$ have a dual 
link $l$ in common and $I_k(x)$ is the modified Bessel function. 

The case of the $SU(N)$ model is slightly more difficult, because the Jacobian 
depends on $\omega(x)$ and the effective action becomes 
\begin{equation} 
\beta \sum_{x,n} \  \rho(x)\rho(x+e_n) \cos (\omega(x)-\omega(x+e_n)) 
+ i N n \sum_x \omega (x) \ . 
\label{sun_action}
\end{equation}
This action can be interpreted as the effective action of the $Z(N)$ spin model 
with a space-dependent coupling constant. Performing the duality 
transformations in this case one finds 
\begin{eqnarray}
\label{sun_dual}
&&Z_{SU(N)}(\beta , N) = \sum_{\{ r(x) \} =-\infty}^{\infty} \ \sum_{\{ n(l) \} 
=-\infty}^{\infty} \ \int_0^1 \prod_p \rho(p) d\rho(p)\nonumber \\
&&\times \prod_p \Phi_{n(p)}(\rho(p)) e^{\lambda N^2 \  \sum_{l} \ (\rho(p)\rho(p^{\prime})^2}\nonumber \\
&&\times \prod_l \ I_{r(x)-r(x+e_n)+N n(l)}(\rho(p)\rho(p^{\prime}) \beta) \ ,
\end{eqnarray} 
where 
\begin{equation} 
n(p) \ = \ n(l_1) + n(l_2) - n(l_3) - n(l_4) \ . 
\label{np}
\end{equation}
Here $l_i, i=1,2,3,4$ are 4 links which form the plaquette $p$. 
In the limit $N\to\infty$ only terms with $n(l)=0$ contribute to the 
$SU(N)$ partition function. Therefore, when $\lambda=0$, the $SU(N)$ partition 
function coincides with the $U(N)$ partition function.

\subsubsection{Jacobian for $U(N)$} 

The integral on the right-hand side of~(\ref{Jnst}) vanishes for all $N$ 
if $\rho>1$. For $U(N)$ the Jacobian is 
\begin{equation}
\Sigma_{U(N)}(\rho)  \ = \  N^2 \ \int_0^{\infty} r \ J_0(N\rho r) \ \mbox{det} 
\, J_{i-j}(r) \  dr  \ .
\label{JUN}
\end{equation}
For $N=1,2$ the integral can be computed exactly 
\begin{eqnarray} 
\label{JU1}
\Sigma_{U(1)}(\rho) \ &=& \ \delta \left ( 1-\rho^2 \right )  \ ,   \\
\Sigma_{U(2)}(\rho) \ &=& \ \frac{4}{\pi \rho} \ \left(1 -\rho^2 \right)^{1/2} \ .
\label{JU2}
\end{eqnarray}
When $N\to\infty$ the integrand in~(\ref{JUN}) is strongly fluctuating and the 
asymptotic expansion can be computed by expanding the determinant at small $r$,
\begin{equation} 
\mbox{det} \ J_{i-j}(r) \ \approx \ \exp \left [ -\frac{1}{4} \ r^2 + A_N r^{2N+2}
\right  ] \ ,  
\label{det_un}
\end{equation}
where 
\begin{equation} 
A_N \ = \ \frac{(-1)^N}{4^{N+1} \ ((N+1)!)^2} \ . 
\label{AN}
\end{equation} 
Treating the second term in the exponent perturbatively, we find the large-$N$ 
asymptotic expansion as  
\begin{equation}
\Sigma_{U(N)}(\rho)  \ = \ 2N^2 \ \exp\left [ -N^2 \rho^2 +  
\frac{(-1)^N}{(N+1)!} \ L_{N+1}(N^2\rho^2) \right ] \ ,
\label{JUN_as}
\end{equation}
where $L_k(x)$ is the Laguerre polynomial. 

\subsection{Mean field and RG for $U(N)$}

To investigate dual models we adopt the procedure developed long ago for the 
$XY$ and vector $Z(N)$ spin models. Namely, we replace the Boltzmann weight 
with its asymptotics when $\beta$ is large enough. 
For the $XY$ model this approximation is known as the Villain representation 
of the $XY$ model. Indeed, in this case we can expect that most configurations 
of the $\rho(x)$ field contributing to the partition function stay away from 
zero. Moreover, we expect that the mean field treatment of the $\rho(x)$ field 
is a good approximation for the partition function. Thus, the analog of the 
Villain form for $U(N)$ spin model is obtained from the approximation 
\begin{eqnarray}
\label{un_approx}
&&I_{r(x)-r(x+e_n)}(\rho(p)\rho(p^{\prime}) \beta)  = 
I_0(\rho(p)\rho(p^\prime) \beta) \nonumber \\ 
&&\times \exp \left [ -\frac{1}{2\beta \rho(p)\rho(p^{\prime})} \ (r(x)-r(x+e_n))^2  
\right ] \ . 
\end{eqnarray}
We use the Poisson resummation formula to sum over $r(x)$ and the mean field 
approximation in the simplest form, which accounts for the substitution 
$\rho(p)\to \langle \rho(p) \rangle = u$ in the action. Performing the 
integration over $r(x)$, we write down the result in the form 
\begin{eqnarray}
\label{un_villain}
Z_{\rm V}(\beta , N) \ = \ Z_{U(1)} (\beta_{\rm eff}) \ Z_{\rm mf}(\beta,u) \ .
\end{eqnarray}
Here, $Z_{U(1)} (\beta_{\rm eff})$ is the partition function of the $XY$ model 
in the Coulomb gas representation 
\begin{eqnarray}
\label{u1_cg}
&&Z_{U(1)} (\beta_{\rm eff}) = \nonumber \\
&&\sum_{\{ m(x) \}=-\infty}^{\infty} \exp\left[ - \pi^2 \beta_{\rm eff} \sum_{x,x^{\prime}} m(x) G_{x,x^\prime} 
m(x^{\prime})\right],
\end{eqnarray}
where $\beta_{\rm eff}=u^2 \beta$; $Z_{\rm mf}(\beta,u)$ is instead the partition 
function for the $\rho(p)$ field in the mean field approximation, 
\begin{eqnarray}
\label{un_mf}
Z_{\rm mf}(\beta,u) \ = \ e^{L^2 \, W(\beta, u)} \ , 
\end{eqnarray}
\begin{eqnarray}
\label{W_mf}
e^{W(\beta, u)} \ = \ e^{ \ln u} \ \int_{0}^1 \ \rho \ \Sigma(\rho) 
\ I_0^4(\beta u \rho) \, d\rho \ ,
\end{eqnarray}
with $u$ determined from the equation 
\begin{equation} 
u \ = \ \frac{\int_{0}^1 \ \rho^2 \ \Sigma(\rho) \ I_0^4(\beta u \rho) \, d\rho}
  {\int_{0}^1 \ \rho \ \Sigma(\rho) \ I_0^4(\beta u \rho) \, d\rho} \ . 
\label{mf_eq}
\end{equation} 

Strictly speaking, a more consistent approach would be to compute the 
free energy for $u$ including also the free energy arising from the $U(1)$ 
part of the partition function~(\ref{u1_cg}). We have checked that this 
produces only exponentially small corrections to the solution 
of Eq.~(\ref{mf_eq}). Once $u$ is computed and $\beta_{\rm eff}$ is fixed, 
the partition function $Z_{U(1)} (\beta_{\rm eff})$ can be analyzed by using 
a conventional RG. It should be clear that, however complicated 
$\beta_{\rm eff}$ may be as function of $\beta$, it is nevertheless
a growing function of $\beta$. Therefore, in the RG flow one can always 
reach the fixed point of the $XY$ model, $\beta_{\rm eff}=2/\pi$, and derive the 
critical point from the relation $\beta_{\rm eff}\approx 0.74$, where 0.74 is the
approximate critical point of the $XY$ model in the Villain formulation. 
Clearly, critical indices cannot depend on the exact dependence of 
$\beta_{\rm eff}$ on $\beta$ and take the same values as in the $XY$ model, 
{\it i.e.} $\eta = 1/4$ and $\nu = 1/2$. 

\begin{figure}[htb]
  \centering
  \includegraphics[width=0.98\columnwidth, clip]{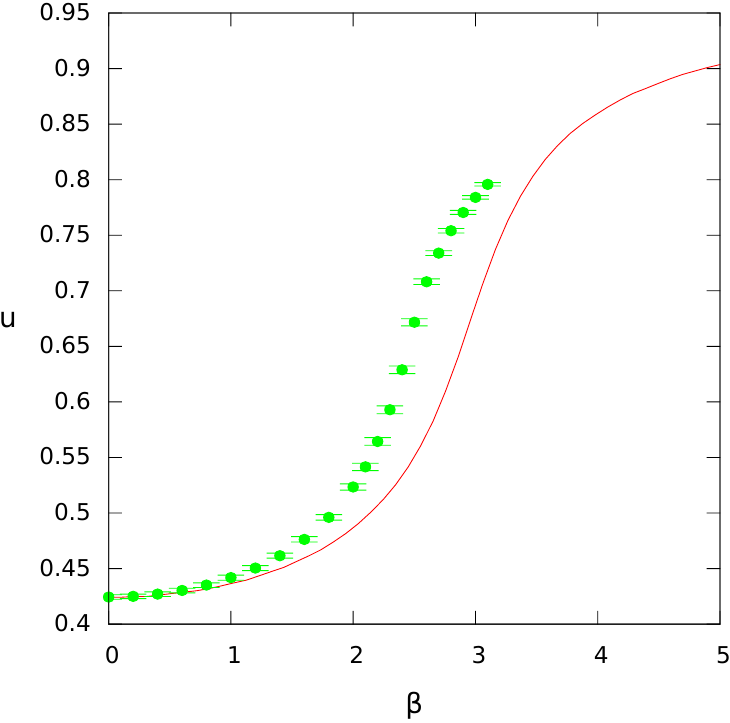}
  \caption{(color online). Dependence on $\beta$ of the solution $u$ of Eq.~(\ref{mf_eq}) for 
  $U(2)$ (solid line) and of the $\rho$ expectation value obtained from 
  Monte Carlo simulations (circles).} 
  \label{fig:u2_rho}
\end{figure}

To justify the use of the mean field approximation, we have numerically solved
Eq.~(\ref{mf_eq}) and calculated $u$ for the $U(2)$ spin model. The resulting 
dependence of $u$ on $\beta$ is shown in \Fig{fig:u2_rho} and is compared
with the expectation values for $\rho(x)$ computed via Monte Carlo simulations 
with the full partition. One can observe a rather good agreement between mean 
field and numerical values over all the range of $\beta$-values studied. 

To perform the mean field study of $U(N)$ models for other values of $N$ one can
calculate the Jacobian~(\ref{JUN}) using numeric integration or the asymptotic 
expansion~(\ref{JUN_as}). In both approaches, we considered several 
values of $N$, namely $N=2,4,5,8$. 
The overall picture appears to be qualitatively similar for all $N$:
for $\beta$ small, $u$ is almost a constant $u \sim {\cal{O}}(N^{-1})$; 
when $\beta$ grows, $u$ starts slowly increasing, with the rate of increase 
reaching its maximum in the region $\beta \sim {\cal{O}}(N^2)$. When $\beta$ 
increases further, $u$ continues slowly moving to the value $u=1$ when 
$\beta\to\infty$. This hints to the fact that the critical point 
of the $U(N)$ model scales as $N^2$, with increasing $N$. 

\section{Details of numerical simulations}  

The observables introduced in this work can be related 
\begin{itemize}[leftmargin=*]

\item 
to {\em full} $U(N)$ and $SU(N)$ spins, 
\begin{equation} 
\chi_{u_N}(x) \ = \ \rho(x) \ e^{i\omega(x)} \ ,
\label{uN_spin}
\end{equation}
\item 
or only to their $U(1)$ part,
\begin{equation} 
\chi_{u_1}(x) \ = \ e^{i\omega(x)} \ .
\label{u1_spin}
\end{equation} 
\end{itemize} 

The corresponding magnetization is defined as follows:
\begin{equation}
\label{complex_magnetization}
M_L^u \ = \  \frac{1}{L^2} \ \sum_x \chi_{u}(x) \ = \ |M_L^u| e^{i \psi} \ ,
\end{equation}
where $u$ can stand either for $u_N$ or $u_1$. Then, all observables 
are defined in a standard way:
\begin{itemize}[leftmargin=*]
\item Mean absolute magnetization, $\mean{|M_L^u|}$,
 
\item Magnetization susceptibility,
\begin{equation}
\chi^{(M_L)} \ = \ L^2 (\mean{|M_L^u|^2} - \mean{|M_L^u|}^2) \ ,
\label{chiML_def}
\end{equation}

\item Binder cumulants,
\begin{equation}
U^{(M_L)} \ =\  1 - \frac{\mean{|M_L^u|^4}}{3 \mean{|M_L^u|^2}^2} 
\label{binder_def} \quad \text{or} \quad 
B_4^{(M_L)} = \frac{\langle\left|M_L^u\right|^4\rangle}{\langle\left|M_L^u\right|^2
\rangle^2}\ ,
\end{equation}

\item Second moment correlation length $\xi_2$,
\begin{eqnarray}
\xi_2 & = & \frac{\sqrt{\frac{\chi}{F}-1}}{2 \sin{\pi/L} } \ ,  \nonumber \\
\chi  & = & \mean{\left| \sum_{x,y} \chi_u(x,y) \right|^2} \ , \nonumber \\
F     & = & \mean{\left| \sum_{x,y} e^{2 \pi i x / L} \chi_u(x,y) \right|^2} \ . 
\label{xi2_def}
\end{eqnarray}

\end{itemize}
For the $SU(N)$ model we also introduce the rotated magnetization 
$M_R = |M_L^{u_N}| \cos{N \psi}$, its mean value and susceptibility.

Simulations for the $U(2)$ model were done using the effective model given 
in Eq.~(\ref{PF_sun_eff}) for $\lambda = 0$ and with the exact Jacobian for 
$U(2)$ model given in Eq.~(\ref{JU2}). 
For the update of the $\rho$ variables a heat-bath algorithm was used, 
while $\omega$ variables were updated using the single-cluster 
Wolff algorithm described in~\cite{wolff}. 
In a typical run, for one choice of the couplings, 50000 measurements 
were taken, with a whole lattice heat-bath sweep and 10 Wolff cluster 
updates between two subsequent measurements.

For $SU(N)$ models the original formulation given in Eq.~(\ref{sunpf}) 
was used. The $SU(N)$ matrices $W(x)$ were parametrized by their eigenvalues 
$e^{i \omega_i}$, since both the partition function of the model and the 
observables can be defined in terms of these eigenvalues. For updating the
$\omega_i$ variables the Metropolis algorithm together with the Wolff cluster 
update  using only $Z(N)$ reflections were adopted. We used lattices with size 
ranging from 8 to 256. In a typical run, for one choice of the couplings, 
25000 measurements were taken, with 10 Metropolis updates of the whole lattice
(with 20 hits per lattice site) and two Wolff cluster updates between two
subsequent measurements. 
 
\section{BKT transition in $U(2)$ spin model} 

\begin{figure*}[htb]
  \centering
  \includegraphics[width=\textwidth,clip]{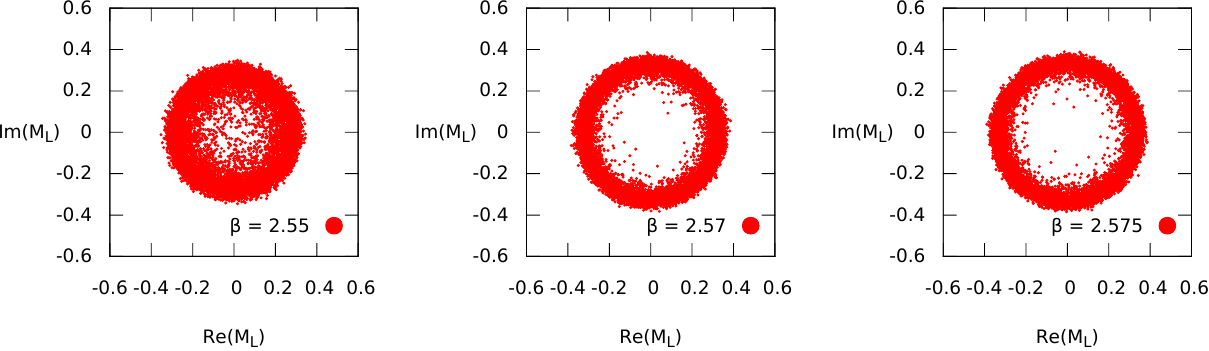}
  \caption{(color online). Scatter plots of the complex magnetization $M_L$ at $\beta = 2.55, 
  2.57, 2.575$ in $U(2)$ on a $512^2$ lattice.}
  \label{fig:scattPlots}
\end{figure*}

In our numerical investigation of the $U(2)$ spin model we proceed in three 
steps: we first check the existence of a BKT transition to a massless phase, 
then locate the critical coupling as precisely as possible and, finally, check
its universality character via the determination of critical indices.

A clear indication of the two-phase structure is provided by scatter plots of
the complex magnetization $M_L$ at different values of $\beta$ and for all
considered lattice sizes. Increasing $\beta$ we observe a transition from a
disordered phase, characterized by a uniform distribution around zero, to a
massless phase, characterized by a ``ring distribution''. Such observation is
reported in~\Fig{fig:scattPlots} for $L=512$.

The next step is the precise location of the phase transition, {\it i.e.} the
identification of the critical coupling $\beta_{\rm c}$. To accomplish this task,
following~\cite{Hasenbusch:2005xm,HasenbuschBinder}, we can make use of
appropriate fit Ans\"atze both for the second-moment correlation length and
for the Binder cumulant {\it versus} the lattice size.
The second-moment correlation length and the Binder cumulant, in correspondence
to a BKT transition, are indeed found to be RG-invariant quantities and to take
universal values that are known for the $XY$ 
model~\cite{Hasenbusch:2005xm,HasenbuschBinder}.
Also their finite size scaling (FSS) behavior has been studied within the 2D 
$XY$ model providing us with the following Ans\"atze:
\begin{itemize}[leftmargin=*]
\item for the second-moment correlation length,
\begin{equation}\label{eq:ksi2Ans}
\frac{\xi_{2}(\beta, L)}{L}=a(\beta)-\frac{b(\beta)}{\mathrm{ln}(L)+c(\beta)}\;,
\end{equation}
where $a_{\rm c} =a(\beta_{\rm c}) = 0.7506912\dots$ and $b_{\rm c} = b(\beta_{\rm c})
= 0.212430\dots$ at the critical point in the $XY$ model;
\item for the Binder cumulant,
\begin{equation}\label{eq:bindAns}
B_4^{(M_L)} (\beta, L) = a(\beta) - \frac{b(\beta)}{\mathrm{ln}(L)+ c(\beta)} +
\frac{d(\beta)}{(\mathrm{ln}(L)+ c(\beta))^2}\;,
\end{equation}
where $a_{\rm c} = a(\beta_{\rm c}) = 1.018192\dots$ and $b_{\rm c} = b(\beta_{\rm c})
= 0.017922\dots$ at the critical point in the $XY$ model, and also subleading 
logarithmic corrections are indicated, since they are found to be important.
\end{itemize}

\begin{figure*}[phtb]
\centering
\includegraphics[width=0.98\columnwidth,clip]{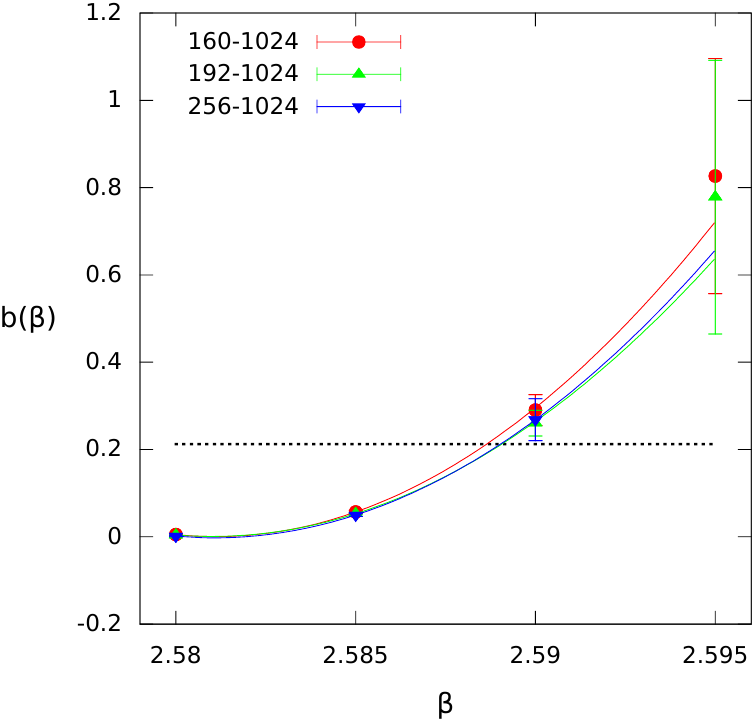}
\hfill
\includegraphics[width=0.98\columnwidth,clip]{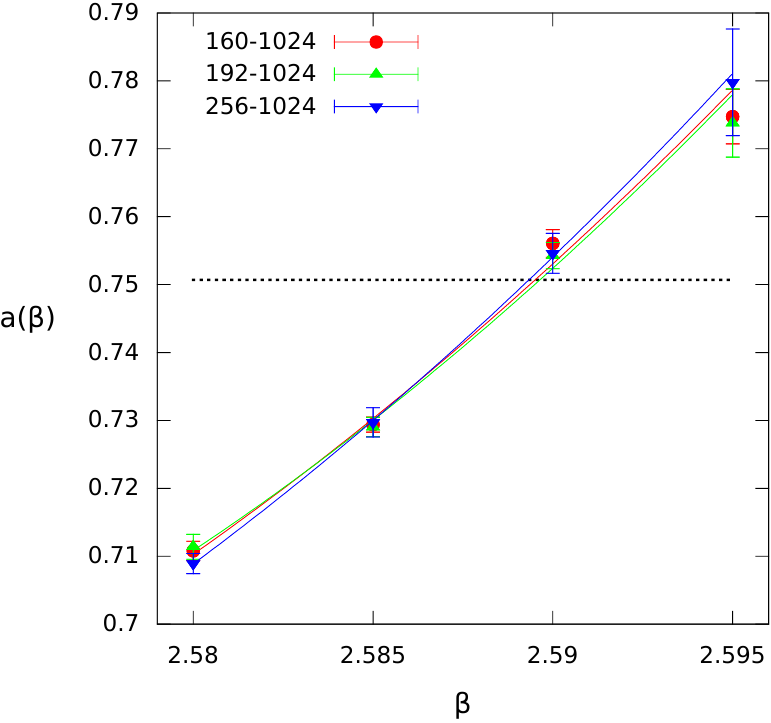}
\caption{(color online). Determination of $\beta_{\rm c}$ from the second-moment correlation 
length.
(Left) Fit parameter $b(\beta)$ extracted from fits of $\xi_{2}$
via~\Eq{eq:ksi2Ans} fixing $a(\beta)$ to the known value $a_{\rm c} = 0.7506912$.
The horizontal dashed line marks the expected value $b_{\rm c}$. 
(Right) Fit parameter $a(\beta)$ extracted from fits of $\xi_{2}$
via~\Eq{eq:ksi2Ans} fixing $b(\beta)$ to the known value $b_{\rm c} = 0.2124300$.
The horizontal dashed line marks the expected value $a_{\rm c}$.}
\label{fig:betacFromKsi2} 
% \end{figure}

% \begin{figure}[phb]
\centering
\includegraphics[width=0.98\columnwidth,clip]{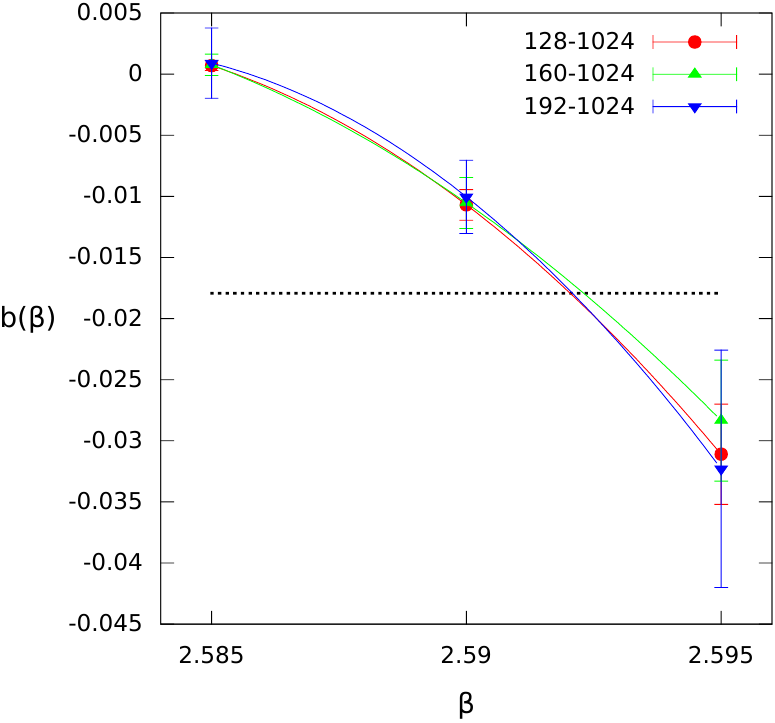}
\hfill
\includegraphics[width=0.98\columnwidth,clip]{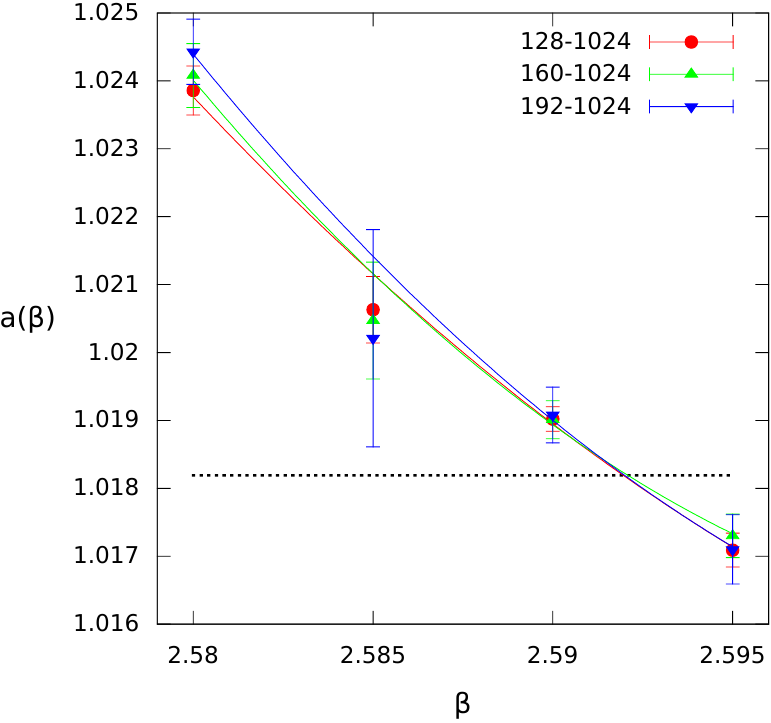}
\caption{(color online). Determination of $\beta_{\rm c}$ from fits of the Binder cumulant $B_4$.
(Left) Fit parameter $b(\beta)$ extracted from fits of $B_4$ via~\Eq{eq:bindAns} 
fixing $a(\beta)$ to the known value $a_{\rm c} = 1.018192$. 
The horizontal dashed line marks the expected value $b_{\rm c}$.
(Right) Fit parameter $a(\beta)$ extracted from fits of $B_4$
via~\Eq{eq:bindAns} fixing $b(\beta)$ to the known value $b_{\rm c} = 0.017922$.
The horizontal dashed line marks the expected value $a_{\rm c}$.}
\label{fig:betacFromBinder}
\end{figure*}

To extract the critical point from the scaling of $\xi_{2}$ and $B_4^{(M_L)}$
with the lattice size $L$, the following methods are used:
\begin{itemize}[leftmargin=*]
\item at each fixed value of $\beta$ the fit of $\xi_{2}$ and $B_4^{(M_L)}$
data is performed, respectively via the Ansatz in~\Eq{eq:ksi2Ans} 
or~\Eq{eq:bindAns}, fixing $b(\beta)$ to the known value $b_{\rm c}$. An 
estimate of $\beta_{\rm c}$ is then given by the $\beta$-value at which 
$a(\beta) = a_{\rm c}$;
\item the same procedure as before, but fixing $a(\beta)=a_{\rm c}$ and 
extracting $\beta_{\rm c}$ from the requirement that $b(\beta_{\rm c}) = b_{\rm c}$.
\end{itemize}
Our results are illustrated in~\Fig{fig:betacFromKsi2} 
and~\Fig{fig:betacFromBinder}, respectively for the second-moment correlation 
length and the Binder cumulant $B_4$. To be conservative, one can quote 
$\beta_{\rm c} = 2.590\pm 0.005$.

As a remark about the fitting procedure, one can observe that its stability,
with respect to the volumes included in the fit, can be inspected by changing
the range of lattice sizes included in the fit and by checking that the
extracted fit parameters are compatible within errors once $L$ is large enough.
We have found that such a stability in the fits is achieved as soon as the
lowest size considered in the fits is $L=128$ for the Binder cumulant
and $L=160$ for the second-moment correlation length.

Once the critical coupling has been estimated, we are able to extract critical
indices out of the FSS behavior at the critical coupling of~\footnote{The symbol
$\beta$ here denotes a critical index and not, as before, the coupling of the
theory.}:
\begin{itemize}[leftmargin=*]
 \item the magnetization $|M_L|$ that is expected to satisfy the scaling law
 \begin{equation}\label{eq:magnFit}
 |M_L| = a L^{-\nicefrac{\beta}{\nu}} \, \mathrm{ln}^r L\;,
 \end{equation}
 \item the susceptibility $\chi^{(M_L)}_L$ that is expected to satisfy the
 scaling law
 \begin{equation}\label{eq:suscFit}
 \chi^{(M_L)}_L = a L^{\nicefrac{\gamma}{\nu}} \, \mathrm{ln}^r L\;.
 \end{equation}
\end{itemize}
Results are summarized in~\tablename~\ref{tab:critIndMagn} 
and~\tablename~\ref{tab:critIndSusc} for the case of magnetization and 
susceptibility, respectively.

\begin{table*}[htb]
\begin{center} 
\caption{Fit parameters $a$, $-\nicefrac{\beta}{\nu}$, and $r$ obtained from 
fits of the magnetization via~\Eq{eq:magnFit} over the size range
$[L_{\mathrm{min}},L_{\mathrm{max}}=1024]$.}
\label{tab:critIndMagn}
\def\arraystretch{1.25}
\begin{tabular}{|@{\quad}c@{\quad}|@{\quad}c@{\quad}|@{\quad}c@{\quad}|@{\quad}c@{\quad}|@{\quad}c@{\quad}|@{\quad}c@{\quad}|}
\hline\hline
$\beta_{\rm c}$ & $L_{\mathrm{min}}$ & $a$ & $-\nicefrac{\beta}{\nu}$ & $r$ & $\chi_r^2$ \\ \hline
% 2.59	&16	&0.72997 $\pm$ 0.00215	&-0.11830 $\pm$ 0.00108	&0.00424 $\pm$ 0.00517	&7.00\\
% 2.59	&32	&0.72275 $\pm$ 0.00298	&-0.12101 $\pm$ 0.00125	&0.01895 $\pm$ 0.00645	&3.90 	\\
2.59	&64	&0.71025 $\pm$ 0.00328	&$-$0.12527 $\pm$ 0.00120	&0.04307 $\pm$ 0.00659	&1.20\\
2.59	&96	&0.70055 $\pm$ 0.00286	&$-$0.12836 $\pm$ 0.00098 &0.06119 $\pm$ 0.00556	&0.35\\
2.59	&128	&0.70089 $\pm$ 0.00469	&$-$0.12826 $\pm$ 0.00151 &0.06055 $\pm$ 0.00879	&0.42\\
2.59	&160	&0.70915 $\pm$ 0.00381	&$-$0.12587 $\pm$ 0.00116 &0.04596 $\pm$ 0.00690	&0.15\\
2.59	&192	&0.70738 $\pm$ 0.00575	&$-$0.12635 $\pm$ 0.00168	&0.04900 $\pm$ 0.01018	&0.18\\
2.59	&256	&0.70322 $\pm$ 0.01075	&$-$0.12746 $\pm$ 0.00299	&0.05600 $\pm$ 0.01854	&0.24\\
\hline\hline 
\end{tabular} 
\end{center}
% \end{table*}

% \begin{table*}[htb]
\begin{center} 
\caption{Fit parameters $a$ and $\nicefrac{\gamma}{\nu}$ obtained from fits of 
the magnetic susceptibility via~\Eq{eq:suscFit} for $r=0$ over the size range
$[L_{\mathrm{min}},L_{\mathrm{max}}=1024]$; including logarithmic corrections with 
$r$ either free or fixed at $r=0.125$ did not improve the fits.}
\label{tab:critIndSusc}
\def\arraystretch{1.25}
\begin{tabular}{|@{\quad}c@{\quad}|@{\quad}c@{\quad}|@{\quad}c@{\quad}|@{\quad}c@{\quad}|@{\quad}c@{\quad}|}
\hline\hline
$\beta_c$ & $L_{\mathrm{min}}$ & $a$ & $\nicefrac{\gamma}{\nu}$  & $\chi_r^2$ \\ \hline
% 2.59 	&16 	&0.00511 $\pm$ 0.00069 	&1.64049 $\pm$ 0.02518 	&226.42\\
% 2.59 	&32 	&0.00403 $\pm$ 0.00034 	&1.68186 $\pm$ 0.01520 	&60.72\\
% 2.59 	&64 	&0.00347 $\pm$ 0.00017 	&1.70759 $\pm$ 0.00869 	&14.24\\
% 2.59 	&96 	&0.00327 $\pm$ 0.00017 	&1.71715 $\pm$ 0.00877 	&10.35\\
% 2.59 	&128 	&0.00308 $\pm$ 0.00014 	&1.72674 $\pm$ 0.00739 	&5.60\\
2.59 	&160 	&0.00293 $\pm$ 0.00010 	&1.73455 $\pm$ 0.00585 	&2.69\\
2.59 	&192 	&0.00282 $\pm$ 0.00010 	&1.74060 $\pm$ 0.00581 	&1.82\\
2.59 	&256 	&0.00273 $\pm$ 0.00013 	&1.74527 $\pm$ 0.00765 	&1.87\\
2.59 	&384 	&0.00249 $\pm$ 0.00005 	&1.75966 $\pm$ 0.00314 	&0.14\\
2.59 	&512 	&0.00241 $\pm$ 0.00003 	&1.76431 $\pm$ 0.00180 	&0.02\\
\hline\hline 
\end{tabular} 
\end{center}
\end{table*}

\section{BKT transition in $SU(N=5,8)$ spin models with adjoint term} 

It is widely expected that the phase transition in $2D$ $SU(N)$ spin
models with only the fundamental representation term is certainly first order
for $N>4$.
On the other hand, if one adds an adjoint-representation term with its own 
coupling $\lambda$, in the limit $\lambda \to \infty$ one gets the partition 
function of $Z(N)$, so there should be two infinite-order phase transitions
in this case. It is possible that the phase structure of $2D$ $Z(N)$ 
spin models is restored already at some finite value of $\lambda$.

\begin{figure*}[phtb]
\includegraphics[width=0.98\columnwidth,clip]{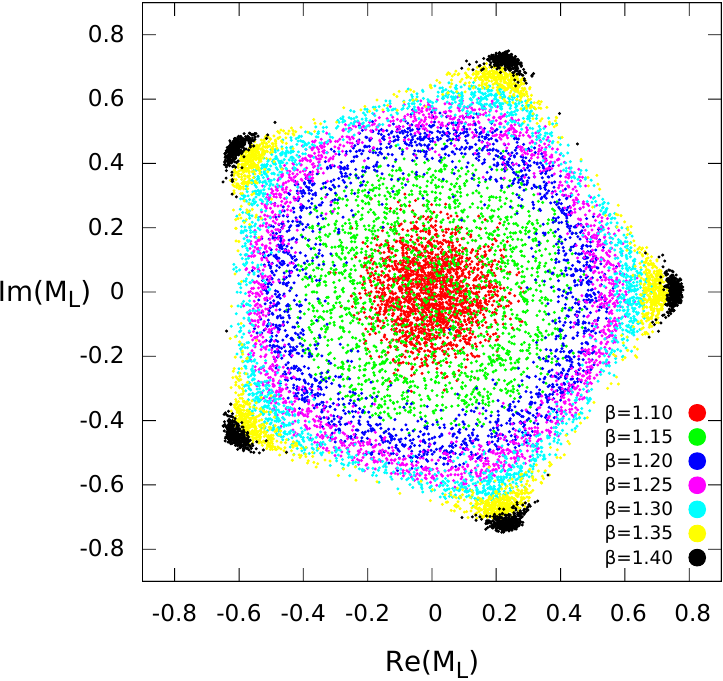}
\hfill
\includegraphics[width=0.98\columnwidth,clip]{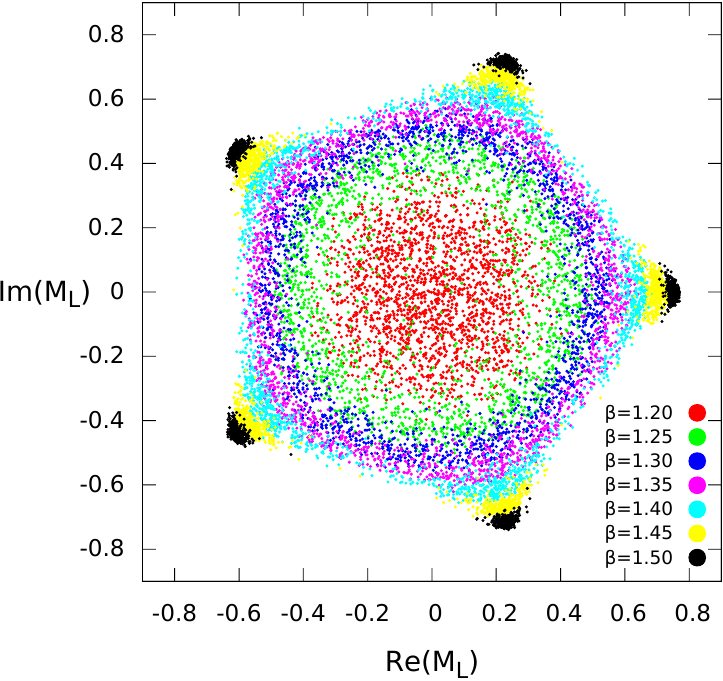}\\

\includegraphics[width=0.98\columnwidth,clip]{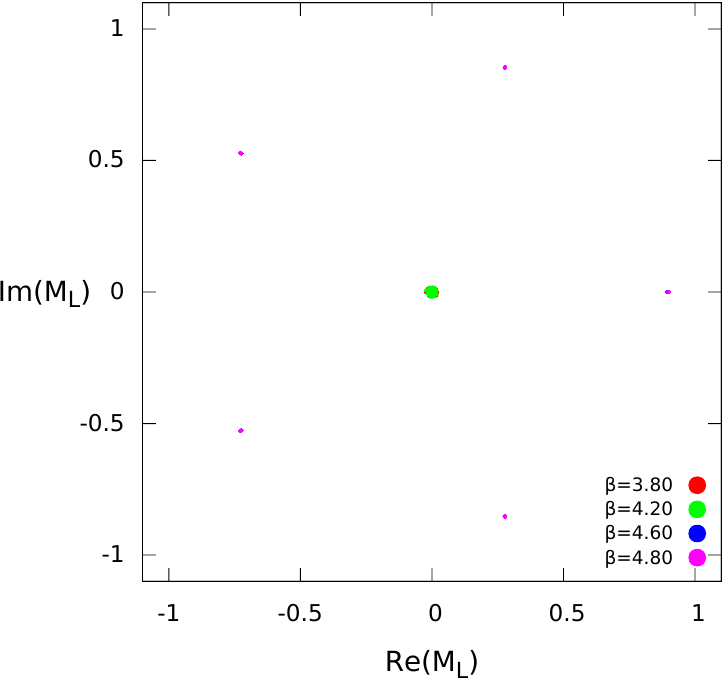}
\hfill
\includegraphics[width=0.98\columnwidth,clip]{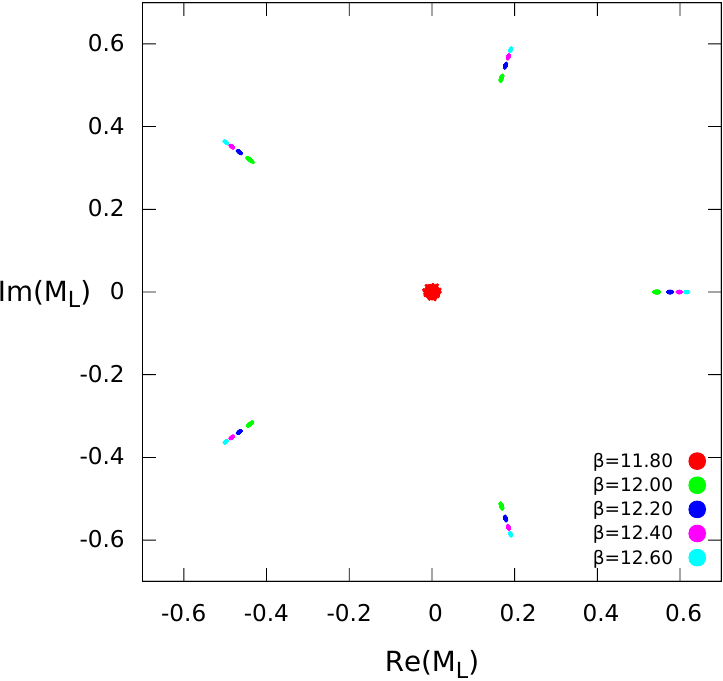}
\caption{(color online). Scatter plots of the complex magnetization $M_L$ in $SU(5)$. (Top-left) $\lambda=1.00$ and $\beta \in [1.10, 1.40]$ on a $128^2$ lattice. (Top-right) $\lambda=0.80$ and $\beta \in [1.20, 1.50]$ on a $128^2$ lattice. (Bottom-left) $\lambda=0.75$ and $\beta \in [3.80, 4.80]$ on a $32^2$ lattice. (Bottom-right) $\lambda=0.00$ and $\beta \in [11.80, 12.60]$ on a $128^2$ lattice.}
\label{fig:su5_scatter}
\end{figure*}

\begin{figure*}[phtb]
\includegraphics[width=0.98\columnwidth,clip]{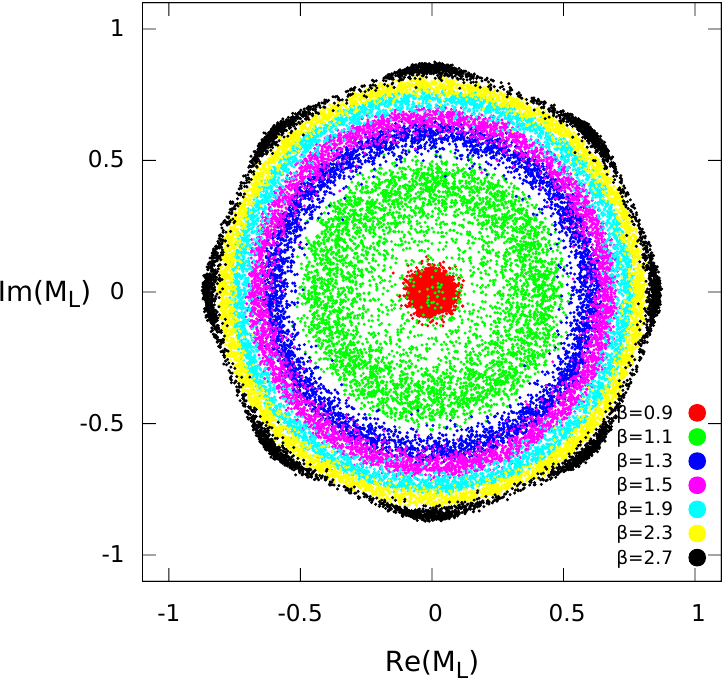}
\hfill
\includegraphics[width=0.98\columnwidth,clip]{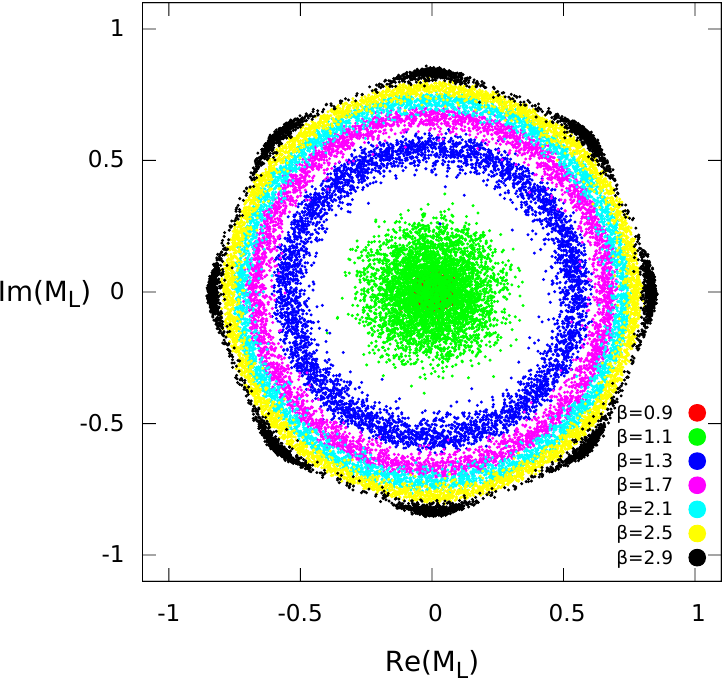}\\
% \vfill
\includegraphics[width=0.98\columnwidth,clip]{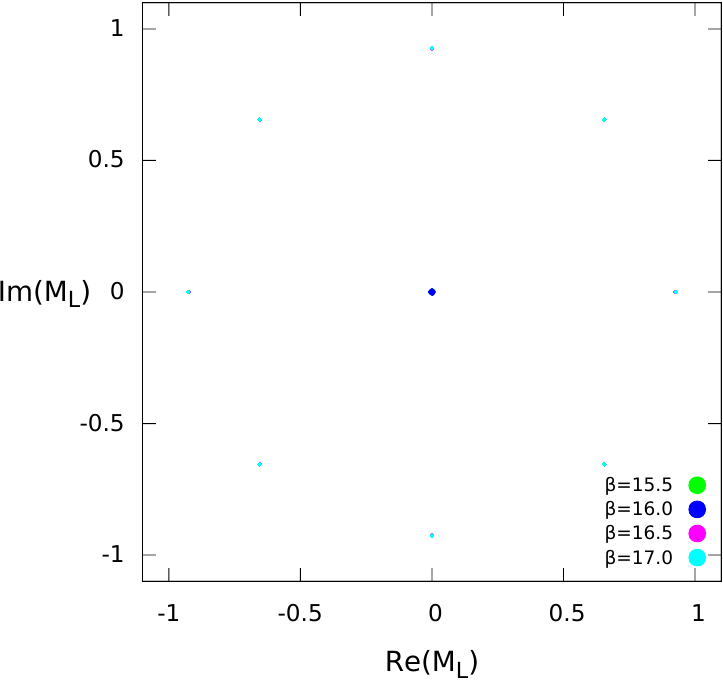}
\hfill
\includegraphics[width=0.98\columnwidth,clip]{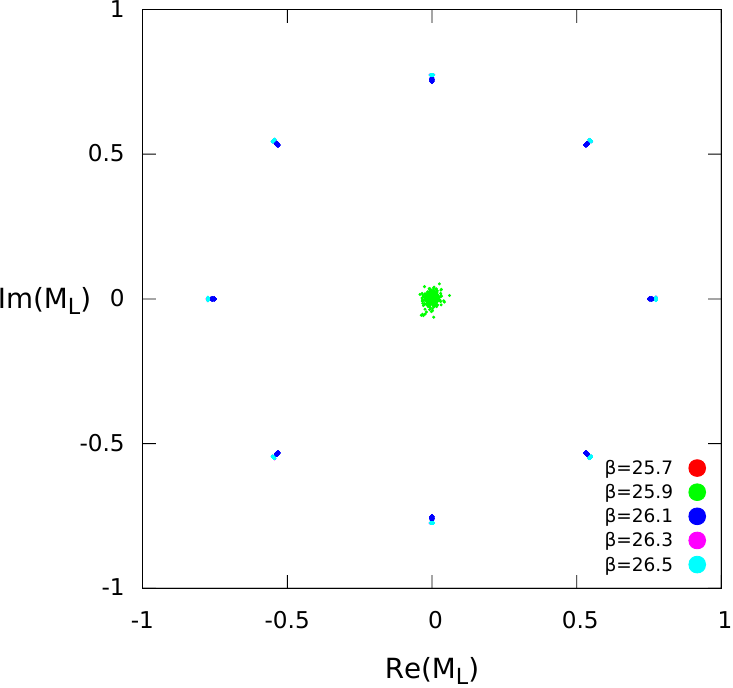}
\caption{(color online). Scatter plots of the complex magnetization $M_L$ in $SU(8)$. (Top-left) $\lambda=2.00$ and $\beta \in [0.90, 2.70]$ on a $128^2$ lattice. (Top-right) $\lambda=1.10$ and $\beta \in [0.90, 2.90]$ on a $128^2$ lattice. (Bottom-left) $\lambda=0.90$ and $\beta \in [15.50, 17.00]$ on a $64^2$ lattice. (Bottom-right) $\lambda=0.00$ and $\beta \in [25.70, 26.50]$ on a $128^2$ lattice.}
\label{fig:su8_scatter}
\end{figure*}

To study the phase transitions in $SU(N)$ models, we have performed 
numerical simulations of $SU(5)$ and $SU(8)$ models. In the $\lambda = 0$ case
our results are consistent with the existence of a first order phase transition.
In particular, we see no intermediate phase with restored $U(1)$ symmetry
and the change of the mean magnetization between phases takes place 
very fast even for small lattice sizes; moreover, around the critical point 
there are signs of a mixed phase, {\it i.e.} the system stays for some time in 
one phase and then jumps to the other one. This behavior continues for small 
$\lambda$ values. However, the phase structure of our model is found to 
significantly depend on the coupling $\lambda$ of the adjoint interaction
term: when $\lambda$ becomes large enough, it changes to that of $2D$ $Z(N)$
model, {\it i.e.} an intermediate phase with $U(1)$ symmetry appears, which 
can be seen from the ring-like scatter plot of the complex magnetization 
(see Figs.~\ref{fig:su5_scatter} and~\ref{fig:su8_scatter}).
A three-phase structure is then found in this case: from the disordered 
low-$\beta$ phase, through a massless phase for $\beta_{\rm c}^{(1)}<\beta
<\beta_{\rm c}^{(2)}$, to an ordered high-$\beta$ phase. There was no attempt 
neither to precisely locate the transition, nor to check its universality 
character.

\begin{figure*}[tb]
\centering
\includegraphics[width=0.48\textwidth,natwidth=127mm,natheight=76mm]{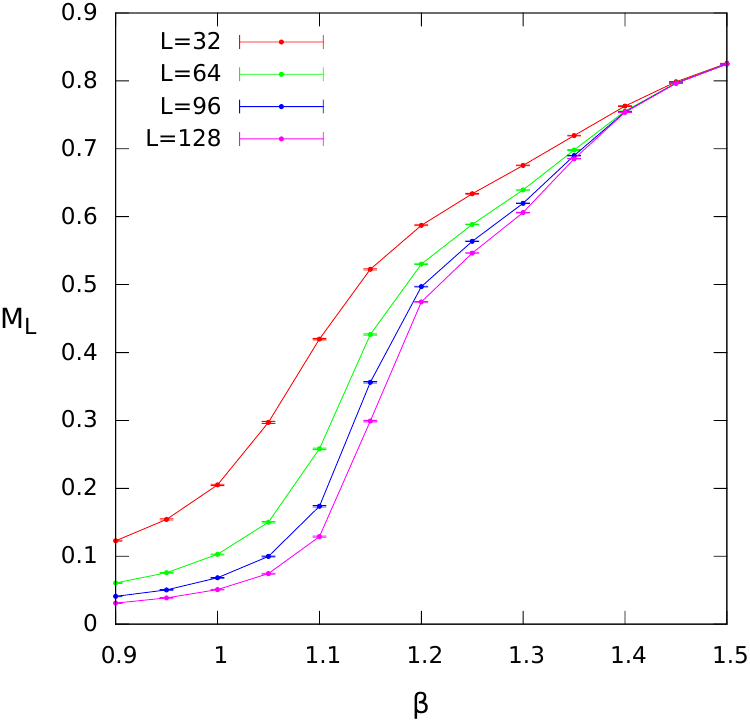}
\hfill
\includegraphics[width=0.48\textwidth,natwidth=127mm,natheight=76mm]{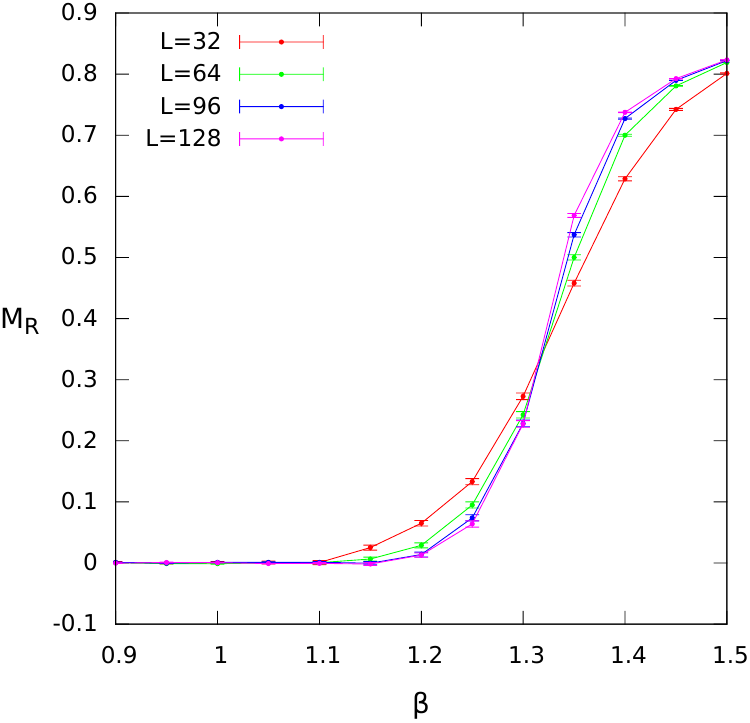} \\
\includegraphics[width=0.48\textwidth,natwidth=127mm,natheight=76mm]{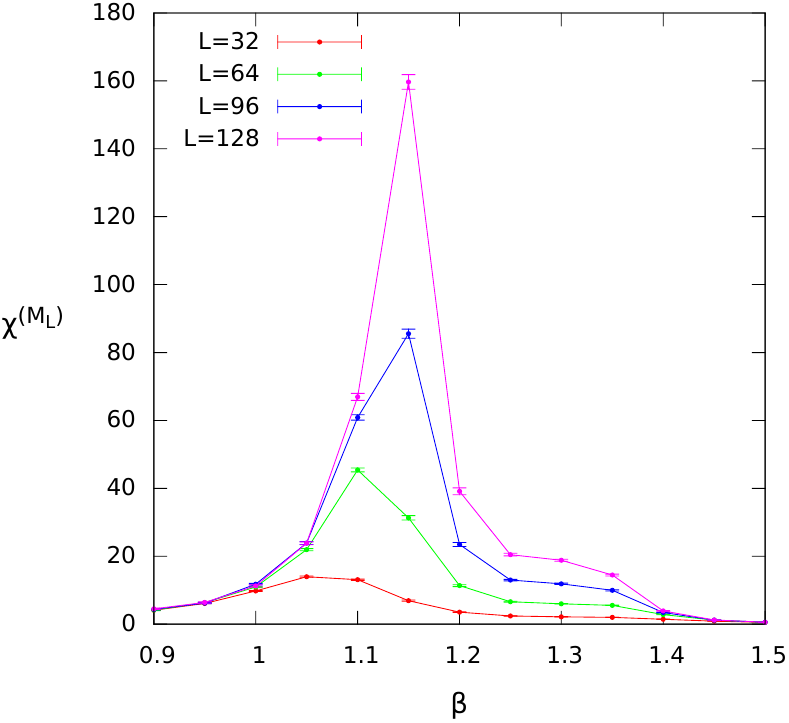}
\hfill
\includegraphics[width=0.48\textwidth,natwidth=127mm,natheight=76mm]{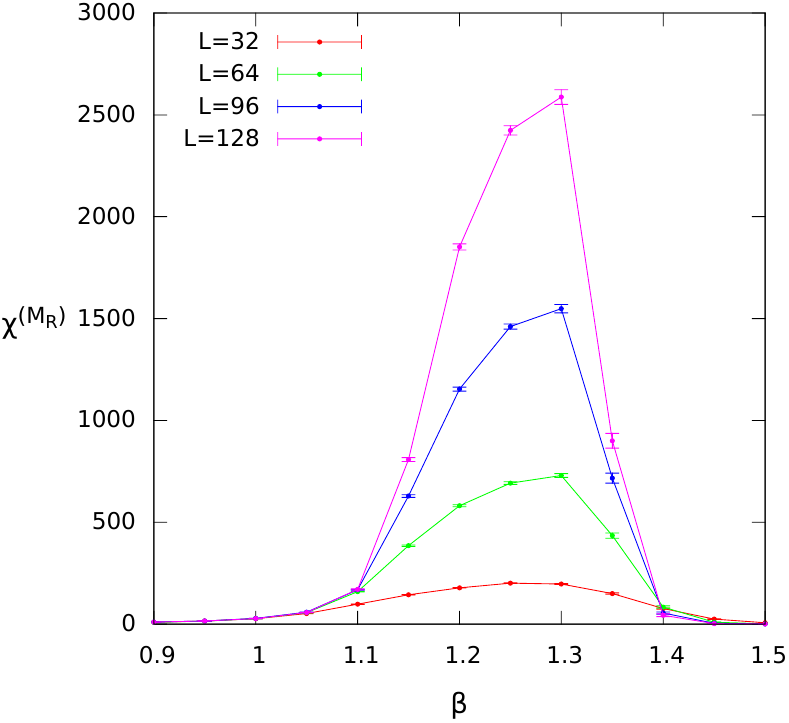}
\caption{(color online). Magnetization (top-left), its susceptibility (bottom-left), rotated 
magnetization (top-right) and its susceptibility (bottom-right) {\it versus} 
$\beta$ in the $SU(5)$ spin model for $\lambda = 1.0$.}
\label{fig:su5_magn}
\end{figure*}

\begin{figure*}[tb]
\centering
\includegraphics[width=0.48\textwidth,natwidth=127mm,natheight=76mm]{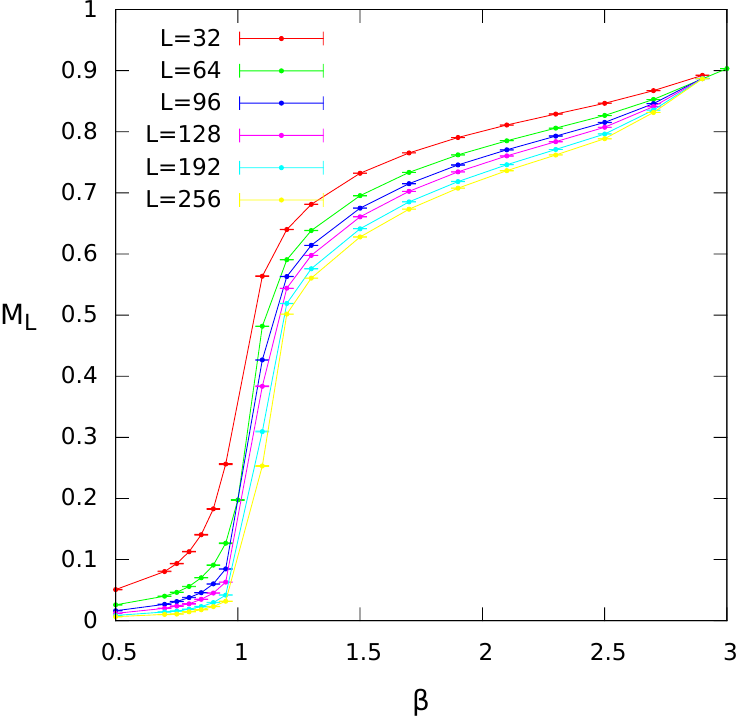}
\hfill
\includegraphics[width=0.48\textwidth,natwidth=127mm,natheight=76mm]{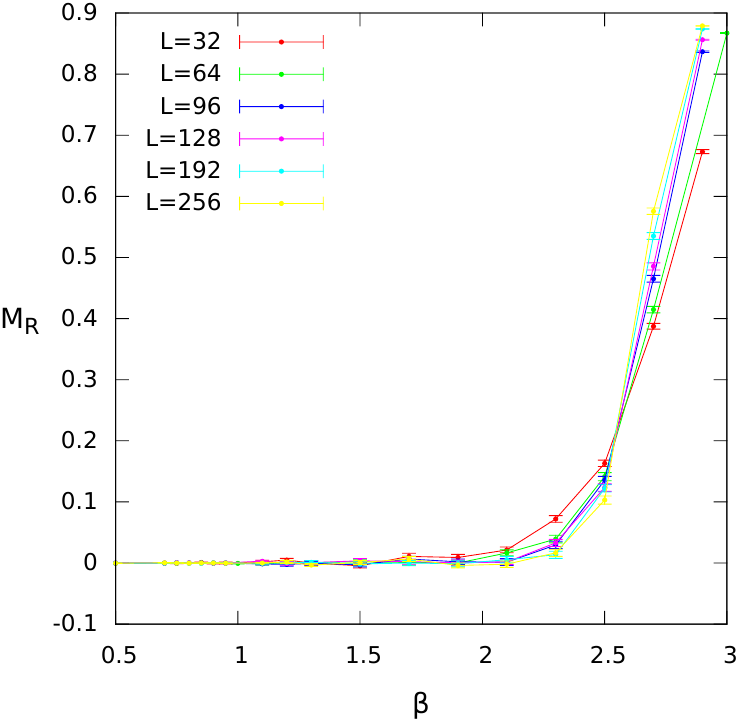} \\
\includegraphics[width=0.48\textwidth,natwidth=127mm,natheight=76mm]{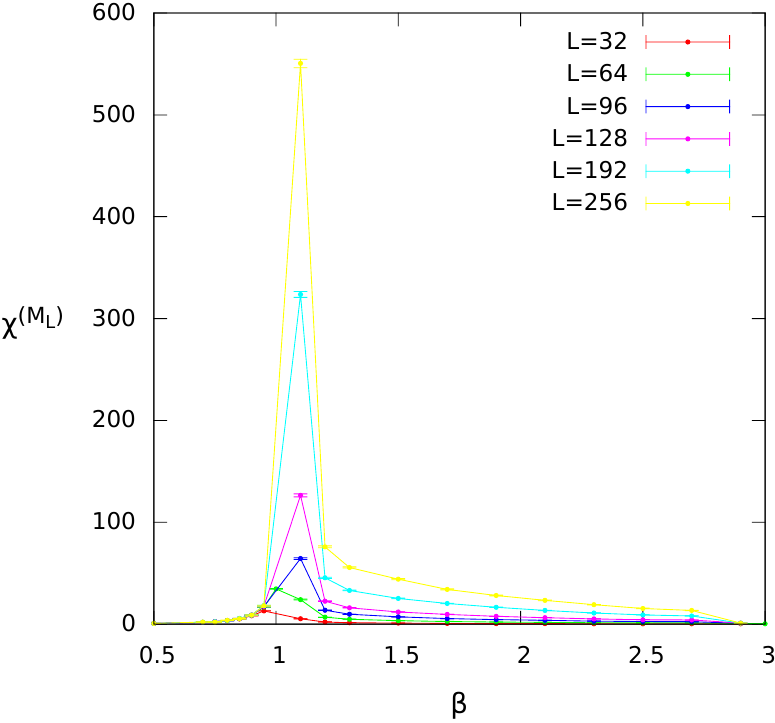}
\hfill
\includegraphics[width=0.48\textwidth,natwidth=127mm,natheight=76mm]{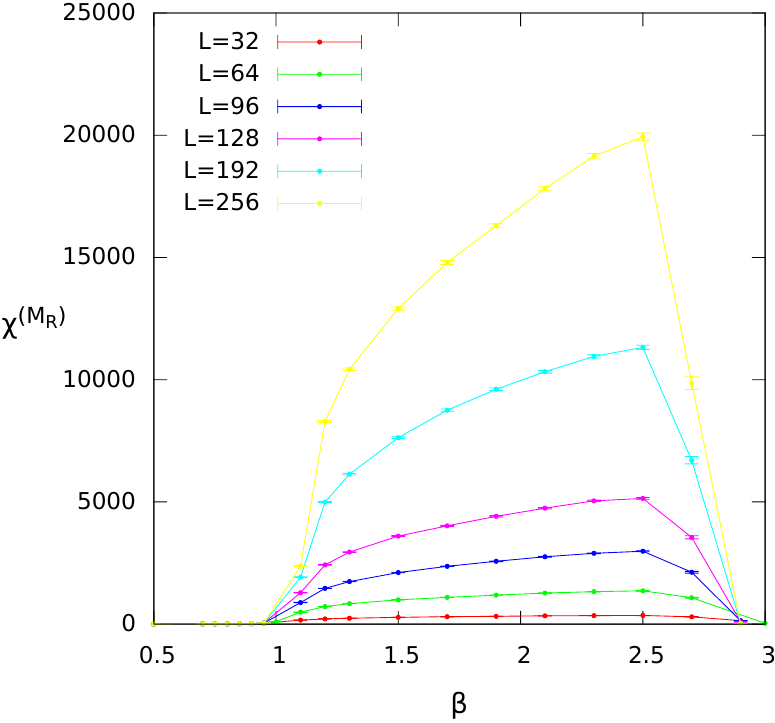}
\caption{(color online). Magnetization (top-left), its susceptibility (bottom-left), 
rotated magnetization (top-right) and its susceptibility (bottom-right) 
{\it versus} $\beta$ in the $SU(8)$ spin model for $\lambda = 2.0$.}
\label{fig:su8_magn}
\end{figure*}

Beyond the scatter plots, further information about the nature of the
observed transitions is provided by the behavior with $\beta$, and for various
fixed $\lambda$ values, of the ordinary and rotated magnetization and their
susceptibilities (\Fig{fig:su5_magn} and~\Fig{fig:su8_magn}).

A remark is in order about the necessity to introduce and measure the rotated
magnetization: what we need is a quantity that is sensitive to the difference
in the magnetization distribution between the massless and ordered phase.
However the two phases differ only for the angular structure of the complex
magnetization, hence the susceptibility and even the Binder cumulant for the
ordinary magnetization are not able to detect the massless to ordered 
transition.
The rotated magnetization, instead, yields a finite value when there is a 
non-trivial angular structure, while it vanishes when $\psi$ is isotropically 
distributed. From the peaks of the susceptibilities of both ordinary and 
rotated magnetization we are able to infer the approximate location of the 
critical points.

The dependence of critical points on $\lambda$ also changes with the kind of 
transition: for small $\lambda$, when the first order phase transition takes
place, $\beta_{\rm c}$ decreases almost linearly with 
$\lambda$; but once we get to the two infinite-order transitions, the
critical points $\beta_{\rm c}^{(1)}$ and $\beta_{\rm c}^{(2)}$ appear 
at much lower values than the ones for $\beta_{\rm c}$ observed during this 
linear decrease.
Whether this means that the transition point starts to decrease faster in the 
vicinity of the critical $\lambda$ or that the transition lines are disjoint, 
remains unclear and requires a more refined study.
It is important, however, to note that, when $\lambda$ further grows, the two 
transition points converge to those of the corresponding $Z(N)$ model.
However, we have currently not yet achieved a precise numerical estimate 
of the minimal $\lambda$ values at which BKT transitions manifest themselves 
in these models.

\section{Summary and Discussion} 

The BKT phase transition occurs in various two-dimensional spin and 
three-dimensional finite-temperature gauge models. All these models have one 
feature in common: they are Abelian models. To the best of our knowledge, this 
type of phase transition has not been observed in any non-Abelian model. 

In this paper we have investigated $U(N)$ and $SU(N)$ spin models in two 
dimensions with the goal of finding evidences for the existence of such phase 
transition. Our findings can be shortly summarized as follows: 

\begin{itemize} 
\item We presented simple arguments, based on the symmetry and 
on Berezinskii-like calculations, that the two-point correlation function in 
$2D$ $U(N)$ models decays with power-law at large $\beta$. Since the 
correlation function has an exponential decay in the strong coupling region, 
this may indicate, similarly to the $XY$ model, a very smooth transition 
from the massive to the massless phase. 

\item More quantitative arguments in a favor of BKT transition have been 
obtained within the effective dual model~(\ref{un_dual}). 
A combination of the mean field and RG analysis led to the conclusion that 
$U(N)$ models belong to the universality class of the $XY$ model. 

\item To verify the above scenario we performed detailed numerical simulations 
of the $U(2)$ model. We have located the critical point of the phase transition 
and computed some critical indices, which appear to agree with those of 
the $XY$ model. 

\item In the case of $SU(N)$ models we have considered a mixed 
fundamental-adjoint action, with two independent coupling constants. 
When the adjoint coupling is small, our numerical results reveal the existence 
of a first order phase transition. When the adjoint coupling grows, the 
fluctuations of $SU(N)$ variables are suppressed and the leading contribution 
comes from the center subgroup. Under such circumstances, one could expect that
the phase structure is similar to that of the $2D$ $Z(N)$ model.  
Indeed, numerical investigations of $SU(N=5,8)$ support the existence of two 
BKT-like phase transitions above a certain critical value of the adjoint 
coupling. 

\end{itemize} 

To clarify completely the physical picture of the phase transitions in these 
models one should still solve some open issues. 
Among them is the extension of the analytical arguments to $SU(N)$ models.  
This can be done with the help of the dual formulation~(\ref{sun_dual}). Then, 
to establish the universality class of $SU(N)$ models and to compute the 
corresponding critical indices, we need to perform large-scale numerical 
simulations of these models. The work on these and some other problems is 
in progress. 

Finally, we would like to stress again that the models studied here can be 
regarded as the simplest ones describing the interaction between Polyakov loops 
in the strong coupling limit of a $3D$ LGT at finite temperature. 
It is well known that in the case of the $SU(N)$ LGT one expects a first 
order phase transition if $N$ is large enough~\cite{teper}. Indeed, our results 
for $N=5,8$, and in the region of small adjoint coupling $\lambda$, agree with 
this conclusion. Of course, the higher-order corrections to the effective 
Polyakov loop model generate all irreducible representations of the group, 
including the adjoint one. At the present stage we can only speculate that the 
values of the corresponding adjoint couplings are rather small and lay below 
the critical $\lambda$ at which the BKT transition occurs. In order to get 
the BKT transition in gauge models, one should add to the Wilson action in the 
fundamental representation the adjoint term. This is an intriguing possibility 
which deserves a further investigation. 

%%%%%%%%%%%%%%%%%%%%%%%%%%%%
% bibliographystyle
\bibliographystyle{apsrev4-1}
\bibliography{./paper_revised.bib}
\end{document}